\documentclass[12pt]{article}
\usepackage{wallpaper}
\usepackage[margin=1in, paperwidth=8.5in, paperheight=11in]{geometry}
\usepackage{amsmath}
\usepackage{graphicx}%
\usepackage{amsfonts}%
\usepackage{amssymb}
\usepackage{color}
\usepackage{float,subfigure}
\usepackage[round]{natbib}
\usepackage{cite}
\usepackage{setspace}
\usepackage{comment}
\usepackage{enumerate}
\usepackage{verbatim}
\usepackage{rotating}
\usepackage{microtype}
\usepackage[normalem]{ulem}
\usepackage{url}
\usepackage[hidelinks]{hyperref} 
\usepackage{cancel}
\usepackage{animate}
\usepackage{mathtools}
\usepackage{epsfig}
\usepackage{epstopdf}

\usepackage{amsthm}

\hypersetup{pdfpagemode=UseNone}
\theoremstyle{plain}
\newtheorem*{thm*}{Theorem} 

\theoremstyle{definition}

\newcommand{\appropto}{\mathrel{\vcenter{
  \offinterlineskip\halign{\hfil$##$\cr
    \propto\cr\noalign{\kern2pt}\sim\cr\noalign{\kern-2pt}}}}}

\newcommand{\bbeta}{ \ensuremath{\boldsymbol{\beta}}}

\newcommand{\btheta}{ \mbox{\boldmath $ \theta $} }
\newcommand{\sig}{ \ensuremath{\sigma}}

\newcommand{\blambda}{ \mbox{\boldmath $\lambda$} }

\newcommand{\bx}{ {\bf x} }

\newcommand{\bz}{ {\bf z} }

\newcommand{\Var}{\mbox{Var}}

\newcommand{\given}{\,\vert\,}

\newcommand{\tbeta}{\mbox{$\text{Beta}$}}

\newcommand{\IG}{\mbox{$\text{IG}$}}

\newcommand{\Pois}{\mbox{$\text{Pois}$}}

\newcommand{\Gam}{\mbox{$\text{Gamma}$}}
\newcommand{\N}{\mbox{Norm}}

\newcommand{\LN}{\mbox{LogNorm}}

\newcommand{\Unif}{\mbox{Unif}}

\begin{document}


\thispagestyle{empty}
\setcounter{page}{0}
\singlespacing
\begin{center}
{\Large \textbf{Evaluating the Informativeness of the Besag-York-Molli\'e CAR Model}} %

\bigskip

\textbf{Harrison Quick$^{1*}$, Guangzi Song$^{1}$, and Loni Philip Tabb$^1$}\\ 
$^{1}$ Department of Epidemiology and Biostatistics, Drexel University, Philadelphia, PA 19104\\
$^{*}$ \emph{email:} hsq23@drexel.edu

\end{center}

\textsc{Summary.}
The use of the conditional autoregressive framework proposed by Besag, York, and Molli\'e (1991; BYM) is ubiquitous in Bayesian disease mapping and spatial epidemiology.  While it is understood that Bayesian inference is based on a combination of the information contained in the data and the information contributed by the model, \emph{quantifying} the contribution of the model relative to the information in the data is often non-trivial.  Here, we provide a measure of the contribution of the BYM framework by first considering the simple Poisson-gamma setting in which quantifying the prior's contribution is quite clear.  We then propose a relationship between gamma and lognormal priors that we then extend to cover the framework proposed by BYM.  Following a brief simulation study in which we illustrate the accuracy of our lognormal approximation of the gamma prior, we analyze a dataset comprised of county-level heart disease-related death data across the United States.  In addition to demonstrating the potential for the BYM framework to correspond to a highly informative prior specification, we also illustrate the sensitivity of death rate estimates to changes in the informativeness of the BYM framework.


\newpage

\doublespacing
\section{Introduction}
The conditional autoregressive (CAR) model popularized by \citet{bym} (BYM) has become ubiquitous in spatial epidemiology and disease mapping.  In addition to being used across a wide range of applications, extensions have been made to spatiotemporal \citep{waller:carlin} and general multivariate \citep{gelfand:mcar,b-r:2015} settings. Missing from the literature, however, is a convenient way to quantify the \emph{informativeness} of the BYM framework
akin to the concept of
``effective sample size'' in the Bayesian clinical trials literature \citep[e.g.,][]{morita:2008}, perhaps due to the complexity of the conditionally dependent nature of spatial models.


The objective of this paper is simple: to provide guidance for how to measure (or alternatively, \emph{control}) the informativeness of the BYM framework.  We begin by anchoring our framework in the conjugate Poisson-gamma setting where measuring the informativeness of the prior distribution is trivial.  We then propose an approach to obtain the approximate informativeness of a lognormal prior and ultimately the BYM CAR model.  After demonstrating the accuracy of this approximation via simulation, we illustrate the potential for oversmoothing using county-level heart disease-relate death data.

\section{Methods}\label{sec:methods}
When modeling rare event and mortality data, we follow the convention set forth by \citet{brillinger} by assuming $y_i \sim \Pois\left(n_i\lambda_i\right)$, where $y_i$ denotes the number of events in region $i$ from a population of size $n_i$ and $\lambda_i$ denotes the underlying event rate, for $i=1,\ldots,I$. Since $\lambda_{i} \sim \Gam\left(a_i,b_i\right)$ is a conjugate prior for the rate parameter in a Poisson likelihood, we can write
\begin{align}
\lambda_{i} \given y_{i}, a_i,b_i \sim \Gam\left(y_{i} + a_i,n_{i}+b_i\right),\label{eq:poisgam}
\end{align}
which yields the interpretations of $a_i$ and $b_i$ as the ``prior number of events'' and ``prior sample size'', respectively.

\begin{figure}[t]
    \begin{center}
        \includegraphics[width=.55\textwidth]{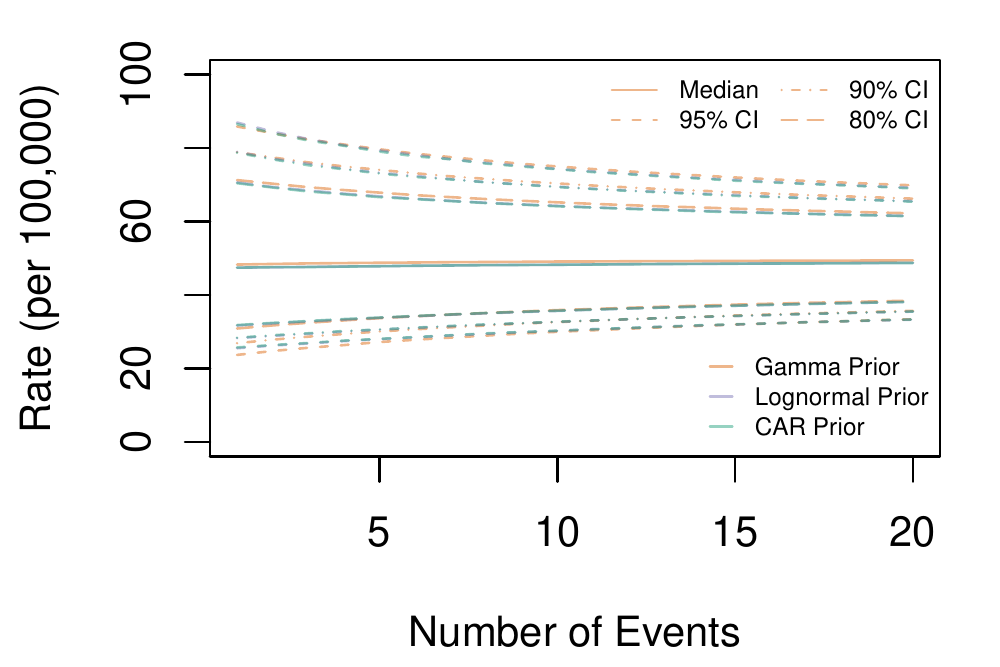}
    \end{center}
    \caption{Comparison of quantiles of the posterior distributions for $\lambda_i$ for gamma, independent lognormal, and \citet{bym}-inspired CAR priors.} 
    \label{fig:compare}
\end{figure}

While the prior specification used to construct the posterior in~\eqref{eq:poisgam} is convenient for illustrating the effect of prior information, it is more common in the disease mapping literature to consider lognormal prior specifications for $\lambda_i$.  Unfortunately, the use of priors like $\lambda_i \sim \LN\left(\mu_i,\sig_i^2\right)$ leads to posterior distributions of an unknown form,
\begin{align*}
p\left(\lambda_i \given y_i,\mu_i,\sig_i^2\right) &\propto \Pois\left(y_i\given n_i\lambda_i\right) \times \LN\left(\mu_i,\sig_i^2\right)\\
&\propto \exp\left[-n_i\lambda_i\right] \times \exp\left[y_i\log\lambda_i-\frac{\left(\log\lambda_i - \mu_i\right)^2}{2\sig_i^2}\right]
\end{align*}
obfuscating the effect of prior information on the posterior distribution.  Thus, to better elucidate the effect of prior information when using lognormal priors, we may wish to construct a prior $\lambda_i\sim \LN\left(\mu_i,\sig_i^2\right)$ that contains approximately the same information as $\lambda_i \sim \Gam\left(a_i,b_i\right)$.  To achieve this, a natural choice may be to equate the mean and variance of their respective distributions; i.e.,
\begin{align}
E\left[\lambda_i\given a_i,b_i\right] =& E\left[\lambda_i\given \mu_i,\sig_i^2\right] &\implies&&  a_i\slash b_i &= \exp\left[\mu_i+\sig_i^2\slash2\right]\label{eq:lognormal}\\
V\left[\lambda_i\given a_i,b_i\right] =& V\left[\lambda_i\given \mu_i,\sig_i^2\right] &\implies&&  a_i\slash b_i^2 &= \left(\exp\left[\sig_i^2\right]-1\right) \exp\left[2\mu_i+\sig^2\right].\notag
\end{align}
From these equations, we can then write $\mu_i$ and $\sig_i^2$ as functions of $a_i$ and $b_i$ --- i.e., $\sig_i^2 = \log\left(1\slash a_i +1\right)$ and $\mu_i = \log\left(a_i\slash b_i\right) - \sig_i^2\slash 2$.
To evaluate the performance of this approximation, Figure~\ref{fig:compare} compares quantiles of the posterior distribution for $\lambda$ given $y$ resulting from a gamma distribution with $a=8.75$ and $y$ taking values $\left\{1,2,\ldots,20\right\}$ with $a\slash b = y\slash n = \lambda_0$ to that resulting from our lognormal approximation, where $\lambda_0$ corresponds to a rate of 50 events per 100,000. 
Based on these results, we claim that the prior $\lambda_i \sim \LN\left(\mu_i,\sig_i^2\right)$ is approximately as informative as $\lambda_i \sim \Gam\left(a_i,b_i\right)$ when we define $\mu_i$ and $\sig_i^2$ in this way; further support for this claim is provided via simulation in Section~3.

While~\eqref{eq:lognormal} measures the informativeness of independent lognormal prior distributions, spatial models such as the BYM framework utilize \emph{conditionally-dependent} prior distributions.  Specifically, if we employ model structures which explicitly model the correlation between $\lambda_i$ and the remaining $\lambda_{j}$, $j\ne i$, then the informativeness of our model is not dictated by the \emph{marginal} mean and variance of $\lambda_i$, but instead by the \emph{conditional} mean and variance, denoted $E\left[\lambda_i\given \blambda_{(i)}\right]$ and $V\left[\lambda_i\given \blambda_{(i)}\right]$, respectively, where $\blambda_{(i)}$ denotes the vector $\left(\lambda_1,\ldots,\lambda_I\right)^T$ with the $i$th element removed. In the context of the CAR model of \citet{bym}, we assume
\begin{align}
\lambda_i \given \bbeta,\bz,\sig^2 \sim \LN\left(\bx_i^T\bbeta + z_i,\sig^2\right), \label{eq:bym}
\end{align}
where $\bx_i$ denotes a $p\times 1$ vector of region-specific covariates with corresponding regression coefficients, $\bbeta$, and $\bz=\left(z_1,\ldots,z_I\right)^T$ denotes a vector of spatial random effects such that
\begin{align}
z_i \given \bz_{(i)},\tau^2 \sim \N\left(\sum_{j\sim i} z_j \slash m_i, \tau^2\slash m_i\right),\label{eq:car}
\end{align}
where $j\sim i$ denotes that regions $i$ and $j$ are neighbors and $m_i$ denotes the number of regions that neighbor region $i$.  As shown in Web Appendix~A, integrating $\bz$ out of~\eqref{eq:bym} leads to a conditional distribution for $\log \lambda_i$ whose precision is bounded below by $1\slash\left(\sig^2+\left[\sig^2+\tau^2\right]\slash m_i\right)$, which we could express in terms of the model's ``informativeness'' as
\begin{align}
\widehat{a}_0 = 1\slash \left(\exp\left[\sig^2+\left(\sig^2+\tau^2\right)\slash m_0\right] - 1\right), \label{eq:inform}
\end{align}
based on the approximation in~\eqref{eq:lognormal} for a baseline number of neighbors, $m_0$. The bound in~\eqref{eq:inform} is achieved when a region neighbors all $I-1$ of the remaining regions, and the precision approaches $1\slash \left(\sig^2+\tau^2\slash m_i\right)$ as the posterior estimates for the neighboring $\lambda_j$ become more precise (e.g., by increasing $m_j$ or $y_j$). As a general rule of thumb, we will evaluate~\eqref{eq:inform} for $m_0=3$ neighbors from this point forward.

To demonstrate the properties of the \citet{bym}-inspired model from~\eqref{eq:bym} and~\eqref{eq:car}, we considered a scenario consisting of $I=50$ regions where each region neighbors all of the remaining $I-1=49$ regions (i.e., $m_i=49$ for all $i$).  We then specified $\sig^2=0.1$ and $\tau^2=0.3$, thus constructing a model with $\Var\left(\theta_i\given \bbeta,\sig^2,\tau^2,\btheta_{(i)}\right)^{-1} = 1\slash\left(\sig^2+\left[\sig^2+\tau^2\right]\slash m_i\right) = 9.25$.  Plugging this into the approximation in~\eqref{eq:lognormal}, we obtain $\widehat{a}_0=8.75$.  As illustrated in Figure~\ref{fig:compare}, this prior specification results in a posterior distribution that is also nearly identical to the posteriors resulting from gamma and independent lognormal priors designed to have the same level of information.  While this scenario --- i.e., $I=50$ regions that all neighbor each other --- is unrealistic, the objective here was simply to illustrate how the expression in~\eqref{eq:inform} can be used to construct priors with the desired properties while avoiding scenarios where spatial models would be inappropriate (e.g., small $I$).

\section{Simulation Study}\label{sec:sim}
While Section~\ref{sec:methods} demonstrates the relationships between the gamma and lognormal prior specifications when the hyperparameters of the lognormal prior specification are fixed and known, we must also demonstrate that these relationships hold when the hyperparameters are \emph{unknown}.  To this end, we conducted a simulation study in which data were generated from a Poisson distribution where the underlying rates were sampled from a gamma distribution --- $y_i \sim \Pois\left(n_i\lambda_i\right)$ and $\lambda_i \sim \Gam\left(a,b\right)$ for $i=1,\ldots, I$, where $a=5$, $b=a\slash\lambda_0$, and $\lambda_0$ corresponds to a rate of 50 events per 100,000 and where $n_i=\text{20,000}$ for all $i$ such that $E\left[y_i\given a,b\right]=10$.  We modeled these data using both the Poisson-gamma and Poisson-lognormal frameworks with all hyperparameters treated as being unknown.  To analyze these data, we compare the following prior specifications:
\begin{align}
\lambda_i &\sim \Gam\left(a,b\right), &a &\sim \Unif\left(0,10\right), &\lambda_0&\sim \Unif\left(0,10^{-3}\right)\label{eq:sim1prior1}\\
\lambda_i &\sim \LN\left(\mu,1\slash \gamma\right), &\mu &\sim \Unif\left(-20,0\right), &\gamma &\sim \Unif\left(0,10\right)\label{eq:sim1prior2},
\end{align}
where $b=a\slash \lambda_0$ and the bounds on the hyperparameters in~\eqref{eq:sim1prior1} and~\eqref{eq:sim1prior2}~are~intended~to~restrict the parameters to a similar range of values (e.g., when $a\approx 10$, $\gamma\approx 10$). The primary goal of this simulation study will be to assess the degree to which the lognormal prior specification in~\eqref{eq:sim1prior2} can produce a posterior distribution similar to that from the prior specification in~\eqref{eq:sim1prior1}.
As the ability to estimate the hyperparameters in~\eqref{eq:sim1prior1} and~\eqref{eq:sim1prior2} depends on the amount of data observed, we let $I=\left\{10,25,50,100,200\right\}$; when $I<200$, multiple sets of data are generated to better assess the models' performance (e.g., 20 sets of data for $I=10$).  All analyses are based on $L=\text{100,000}$ posterior samples obtained using the {\tt rjags} package \citep{rjags} and thinned by a factor of 10 to reduce autocorrelation.

In Figure~\ref{fig:sim1post}, we see that while the informativeness of the gamma prior for $I=200$ is centered around the true value of $a=5$, the lognormal prior yields a slightly less informative posterior.  Results for smaller values of $I$ are provided in Web Appendix~B.  As would be expected, small values of $I$ have much less precision when measuring the informativeness of the priors.

\section{Illustrative Example: Drug-Overdose Death Data}\label{sec:analysis}
We now consider a dataset comprised of the number of heart disease-related deaths (ICD-9: 390--398, 402, 404--429) among those aged 35--54 in 1980 from counties in the contiguous United States.  These data predate the CDC's data confidentiality protections --- namely, that counts less than 10 are suppressed for data dating back to 1989 \citep{cdc:sharing} --- and thus are publicly available without suppression.  And while heart disease was the leading cause of death in 1980, mortality rates in this age bracket were still quite low, resulting in a preponderance of small counts and thus motivating the use of spatial models to produce more reliable estimates.

We first consider a case study using the 77 counties of Oklahoma.  Here, we begin by fitting the standard BYM CAR model based on~\eqref{eq:bym} and~\eqref{eq:car}.  Standard priors were used per \citet{waller:carlin} --- $p\left(\beta_0\right) \propto 1$, $\sig^2 \sim \IG\left(1,1\slash 100\right)$, and $\tau^2 \sim \IG\left(1,1\slash 7\right)$ --- and our MCMC algorithm was run for 50,000 iterations. After fitting the model, we estimate the informativeness of this model, $\widehat{a}_0$, based on~\eqref{eq:inform}. Finally, we refit the model subject to the restriction that $\widehat{a}_0 < 6$ for a county with $m_0=3$ neighbors and explore the implications of this restriction.  We then repeat this same analysis on the remaining 47 states in the contiguous United States --- one state at a time --- minus those with fewer than five counties where the use of a spatial model may not be appropriate.  The goal of this second set of analyses is to highlight the heterogeneity in the informativeness of the CAR model of \citet{bym} when analyzing the same outcome (heart disease-related deaths) at the same spatial scale (counties) from different locations (states).

\subsection{Case study: Heart disease-related deaths in Oklahoma}
While the heart disease-related death rate for those 35--54 in the state of Oklahoma (106.5 deaths per 100,000) was on par with the national average (108.4), Oklahoma's large number of rural counties led to 64 of the 77 counties experiencing fewer than 10 deaths in this age bracket.  This leads to the inferential dilemma which motivates this work --- i.e., we want to use models that \emph{leverage} spatial structure to produce more reliable estimates, but we do not want those models to \emph{overwhelm} the information contained in the data.

\begin{figure}[t]
    \begin{center}
        \subfigure[Simulation Study]{\includegraphics[width=.45\textwidth]{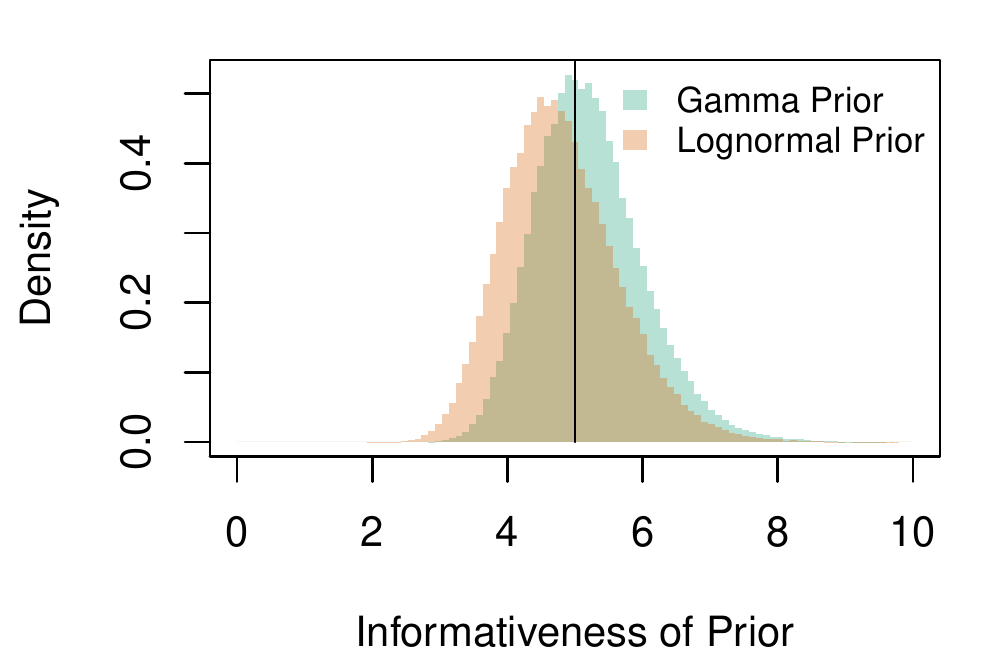}\label{fig:sim1post}}
        \subfigure[Heart Disease in Oklahoma]{\includegraphics[width=.45\textwidth]{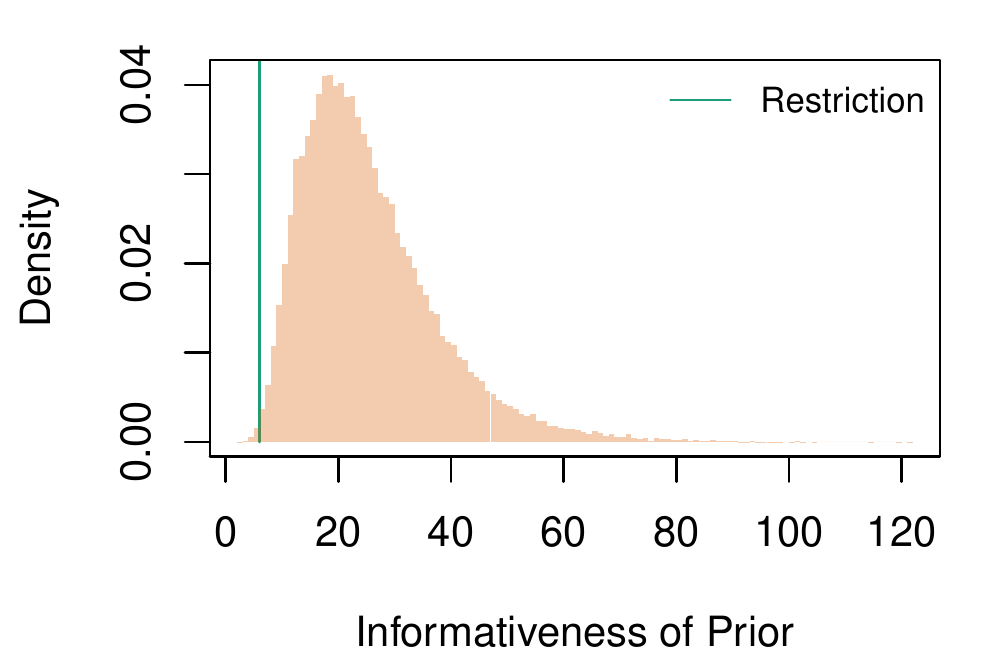}\label{fig:resinfo}}
    \end{center}
    \caption{Posterior distributions of the models' informativeness.  Panel~(a) corresponds to the simulation study in Section~\ref{sec:sim} for $I=200$ compared to the true value of $a=5$, while Panel~(b) corresponds to the case study using death data from Oklahoma compared to the restriction that $\widehat{a}_0 < 6$ for a county with $m_0=3$ neighbors.}
    \label{fig:info}
\end{figure}

\begin{figure}[t]
    \begin{center}
        \subfigure[Rates from Unrestricted Analysis]{\includegraphics[width=.45\textwidth]{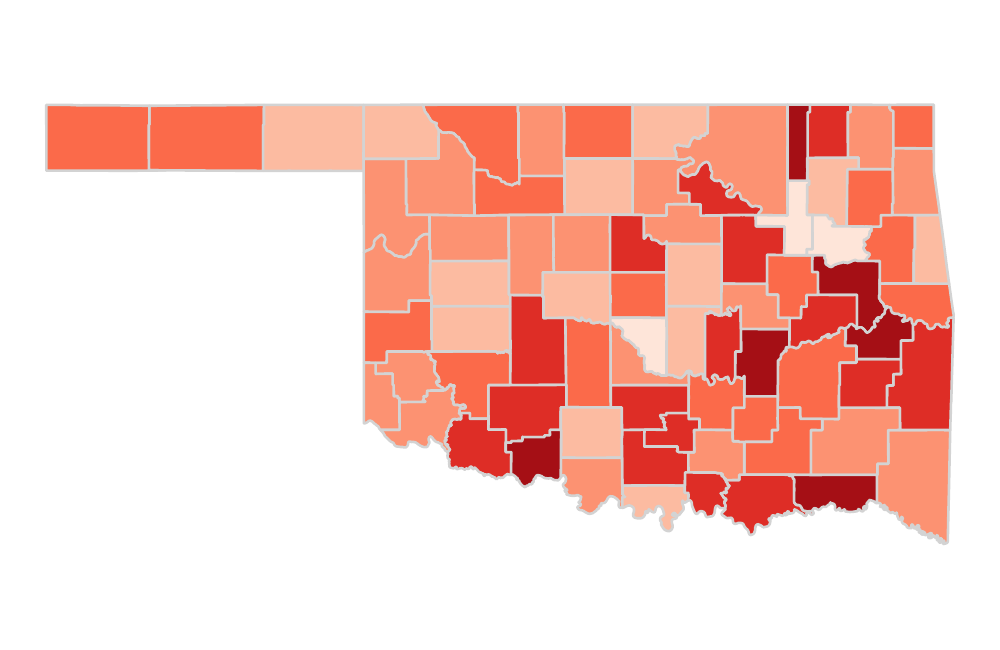}\label{fig:resrate1}}
        \subfigure[Rates from Restricted Analysis]{\includegraphics[width=.45\textwidth]{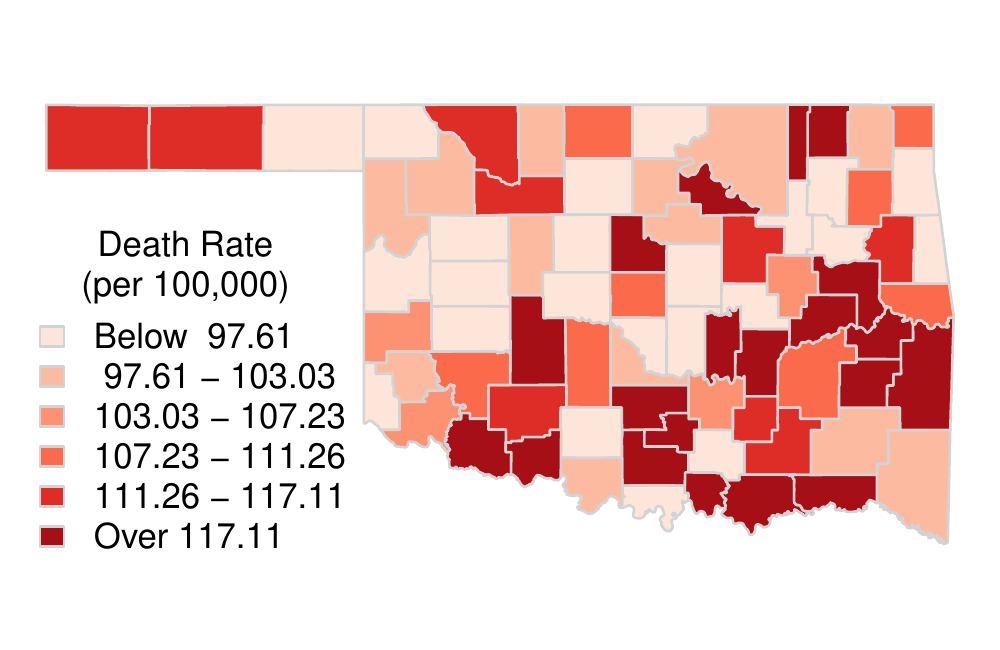}\label{fig:resrate2}}
    \end{center}
    \caption{Posterior medians for the heart disease-related death rates from the unrestricted and restricted analyses, respectively.}
    \label{fig:resrates}
\end{figure}

We begin by analyzing the data using the CAR-based model in~\eqref{eq:bym} without restrictions on the informativeness of the prior specification.  Using the expression for $\widehat{a}_0$ in~\eqref{eq:inform}, the prior specification for this model is approximately equivalent to an additional 23 deaths for a county with $m_0=3$ neighbors, as indicated in Figure~\ref{fig:resinfo}.  To see the effect of such strong prior information, we consider the rate estimates in Figure~\ref{fig:resrates}.  In Figure~\ref{fig:resrate1}, we see that the conventional, unrestricted CAR model yields a relatively smooth map of rates, where the rates in more rural parts of the state resemble those in the urban centers of Tulsa and Oklahoma City. In contrast, if we restrict $\sig^2$ and $\tau^2$ such that the model in~\eqref{eq:bym} contributes fewer than $6$ deaths to our estimates, we obtain a less spatially smooth map, thereby allowing counties with high observed rates to differentiate themselves from their neighbors.  

\subsection{Illustration of heterogeneity in informativeness}
\begin{figure}[t]
    \begin{center}
        \subfigure[Informativeness of the CAR Model]{\includegraphics[width=.45\textwidth]{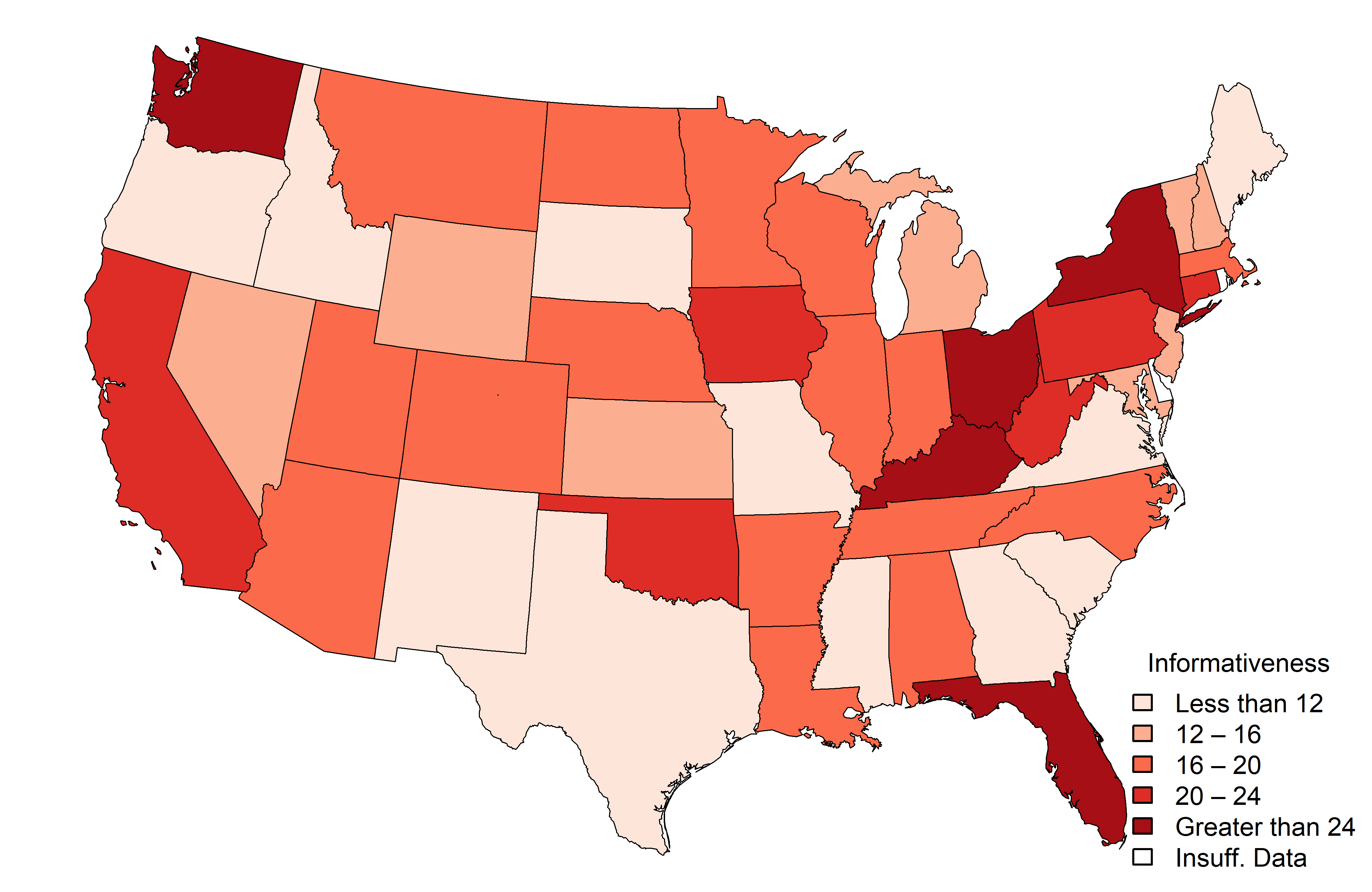}\label{fig:stateinfo}}
        \subfigure[Unrestricted vs.\ Restricted]{\includegraphics[width=.45\textwidth]{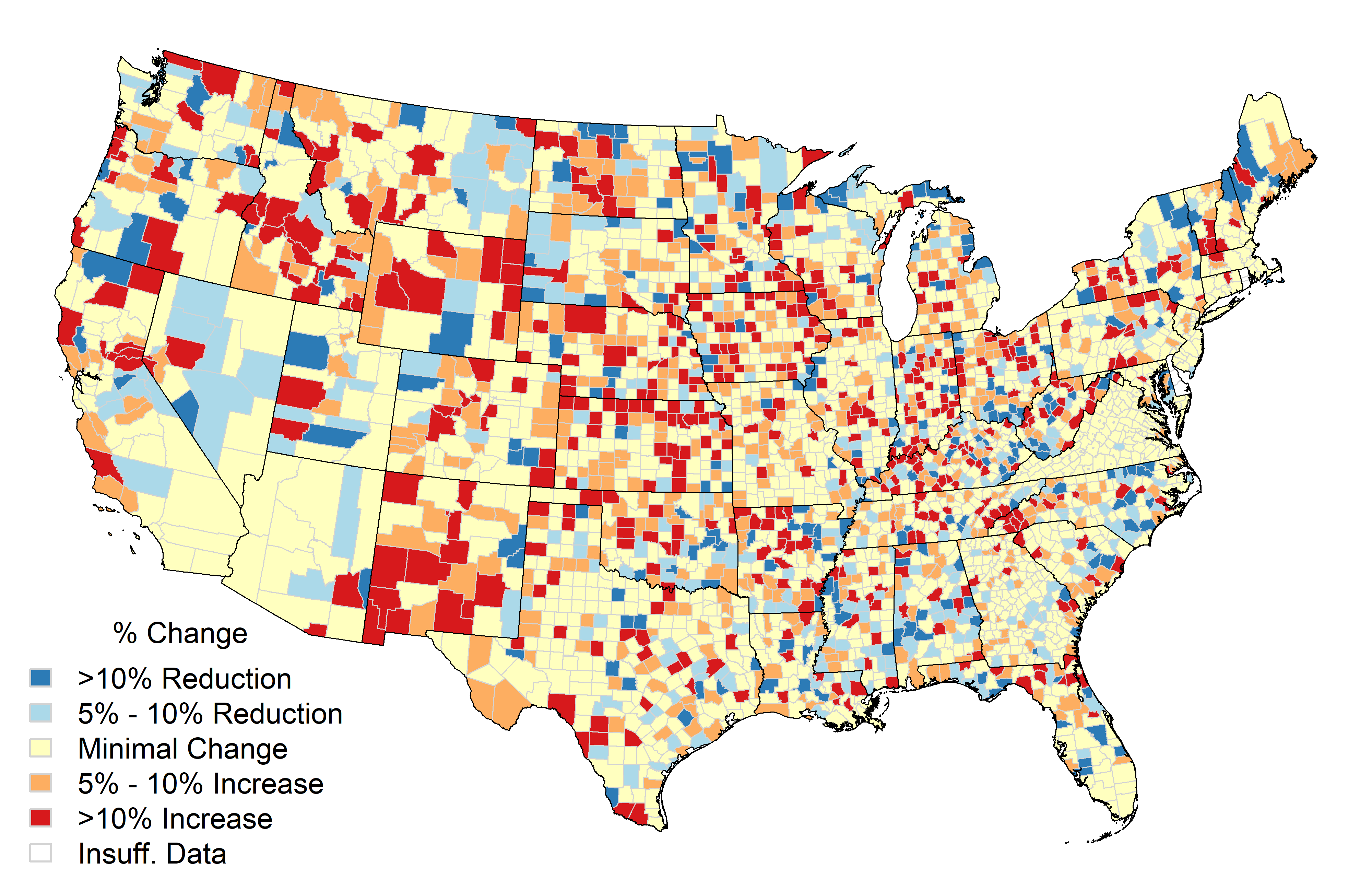}\label{fig:rateratios}}
    \end{center}
    \caption{Results from a state-by-state analysis of the heart disease-related death rates.  Panel~(a) displays the estimated informativeness of the \citet{bym} model within each state, while Panel~(b) displays the percent change resulting from the use of an unrestricted prior specification compared to the restricted (i.e., an ``increase'' indicates that the unrestricted analysis yields higher rates than one whose informativeness is restricted).}
    \label{fig:states}
\end{figure}

We now repeat the above unrestricted analysis on the remaining 47 states in the contiguous United States, minus those with fewer than 5 counties.  Figure~\ref{fig:stateinfo} displays the estimated informativeness measure from~\eqref{eq:inform} for a county with $m_0=3$ neighbors from each state.  Here, we see that the informativeness of the model in~\eqref{eq:bym} varies wildly, ranging from contributing the effect of under 5.5 additional deaths per county in Virginia to over 36 in Ohio. While not shown here, there does not appear to be a discernible pattern between our measure of the state-specific measures of model informativeness, $\widehat{a}_0$, and their respective event rates or other simple summary statistics (e.g., percent of counties with small counts, percent of rural counties, etc.).  Thus, it can be difficult to predict \emph{a priori} the CAR model's informativeness and the extent to which this can alter point estimates (Figure~\ref{fig:rateratios}) and their precision.

\section{Discussion}\label{sec:disc}
The use of spatial models is often motivated by a desire to leverage the spatial structure in the data to improve the precision of estimates from areas with limited data.  While we consider this a perfectly valid rationale, we believe more care should be taken to ensure that these models do not produce estimates that are more precise and more spatially smooth than the data warrant.  A review of the literature \citep[e.g.,][]{bernardinelli,waller:carlin} suggests the use of relatively noninformative priors for the variance parameters, $\sig^2$ and $\tau^2$, which may lead users to believe that the BYM framework itself will not be overly informative.  As we have illustrated here, this does not appear to be the case.

Furthermore, while much research has been done to construct weakly informative priors \citep{gelman2006} or to theoretically derive prior distributions that penalize complexity \citep{simpson2017}, the contribution of this paper is to quantify how informative the \emph{model} is for certain values of $\sig^2$ and $\tau^2$, regardless of the priors used.  Thus, our objective is not to prescribe which priors should be used for these parameters, but instead to provide guidance regarding their specification or potential restrictions on the range of values they are allowed to take.  Finally, it should be noted that while this work has focused on the CAR model proposed by \citet{bym}, similar methods can (and should) be developed for other popular disease mapping approaches such as the CAR framework of \citet{leroux:car} and the directed acyclic graph autoregressive model of \citet{datta:dagar}.

\bibliographystyle{jasa}
\bibliography{chr_ref}

\end{document}